\documentclass[10pt,twocolumn,letterpaper]{article}

\usepackage[pagenumbers]{cvpr} % To force page numbers, e.g. for an arXiv version

\usepackage[dvipsnames]{xcolor}

\definecolor{cvprblue}{rgb}{0.21,0.49,0.74}
\usepackage[pagebackref,breaklinks,colorlinks,citecolor=cvprblue]{hyperref}

\def\paperID{606} % *** Enter the Paper ID here
\def\confName{CVPR}
\def\confYear{2025}

\title{FFaceNeRF: Few-shot Face Editing in Neural Radiance Fields}

\author{
$\text{Kwan Yun}^1$ \quad  $\text{Chaelin Kim}^1$ \quad $\text{Hangyeul Shin}^2$ \quad  $\text{Junyong Noh}^1$
 \vspace{1mm}
 \\$^1$KAIST, Visual Media Lab \quad $^2$Handong Global University
}

\begin{document}

\setlength{\belowcaptionskip}{-2mm}
\twocolumn[{
\maketitle\vspace{-5mm}
\vspace{-2mm}
\begin{center}
    \begin{tabular}{c}
    \vspace{-1mm}
    \includegraphics[width=0.9\linewidth]{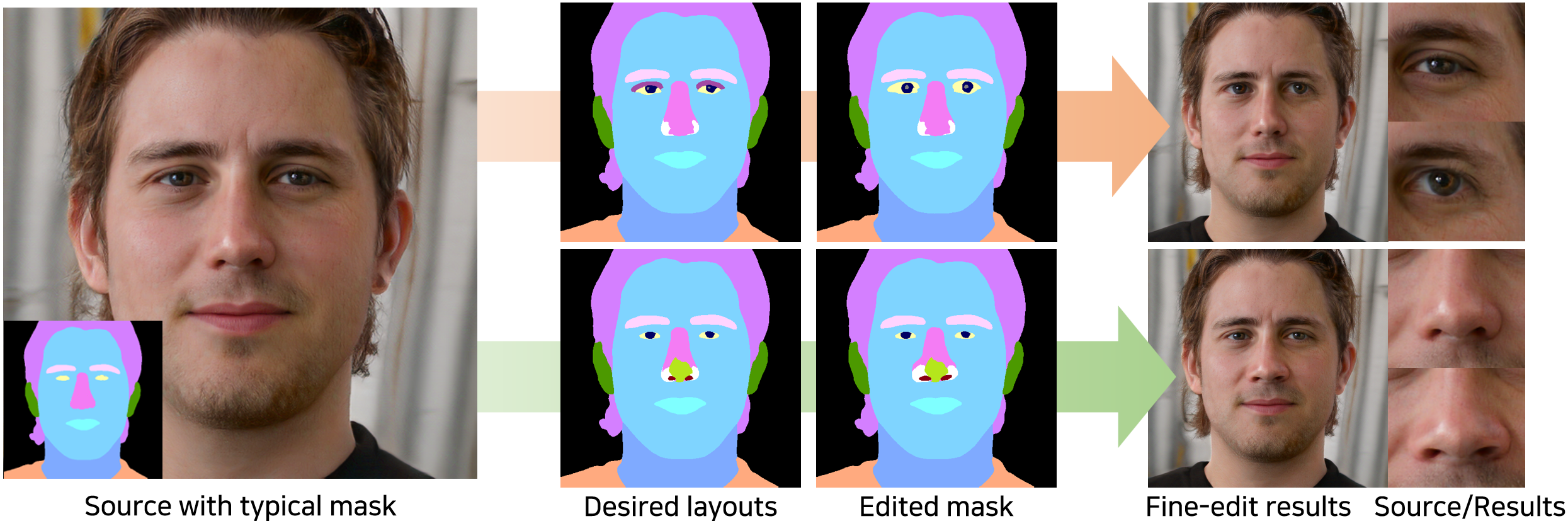}
    \end{tabular}
    \vspace{-1mm}
    \captionof{figure}{Results of FFaceNeRF. With few-shot training, our method can edit 3D-aware images from desired layouts.}
    \label{teaser}
    \vspace{3mm}
\end{center}

}]
%-------------------------------------------------------------------------
\begin{abstract}
Recent 3D face editing methods using masks have produced high-quality edited images by leveraging Neural Radiance Fields (NeRF). Despite their impressive performance, existing methods often provide limited user control due to the use of pre-trained segmentation masks. To utilize masks with a desired layout, an extensive training dataset is required, which is challenging to gather. We present FFaceNeRF, a NeRF-based face editing technique that can overcome the challenge of limited user control due to the use of fixed mask layouts. Our method employs a geometry adapter with feature injection, allowing for effective manipulation of geometry attributes. Additionally, we adopt latent mixing for tri-plane augmentation, which enables training with a few samples. This facilitates rapid model adaptation to desired mask layouts, crucial for applications in fields like personalized medical imaging or creative face editing. Our comparative evaluations demonstrate that FFaceNeRF surpasses existing mask based face editing methods in terms of flexibility, control, and generated image quality, paving the way for future advancements in customized and high-fidelity 3D face editing. The code is available on the {\href{https://kwanyun.github.io/FFaceNeRF_page/}{project-page}}.
\end{abstract}
\vspace{-2mm}
\vspace{-3mm}
\section{Introduction}
In the growing field of digital image synthesis, the ability to create realistic and controllable facial images is crucial, especially for applications in VR/AR, personalized avatar creation, and medical purposes. To deliver high-fidelity facial images and support their extensive customization, precise manipulation of facial attributes and intrinsic features is essential. To achieve this, image-based editing methods have been proposed~\cite{shen2020interfacegan,chen2016infogan,abdal2021styleflow,tewari2020stylerig}. Although these methods successfully achieve high performance in generating edited images by controlled features, they suffer from inconsistency when the camera view is changed.

Recent advancements in 3D face editing, driven by generative Neural Radiance Fields (NeRF)~\cite{mildenhall2021nerf,chan2022efficient}, have significantly improved the resolution and quality of the edited images. These methods enable 3D-aware face generation and editing using segmentation masks. Early studies~\cite{sun2022fenerf,chen2022sofgan} required large training datasets to build 3D segmentation fields, while recent studies~\cite{jiang2022nerffaceediting,sun2022ide}  have reduced the burden of data collection by utilizing pre-trained segmentation networks. Unfortunately, the use of a pretrained segmentation network necessitates a mask with a fixed layout, making it impossible to edit the regions that are not included in the segment labels of the layout. The purposes of editing may differ from session to session, requiring a different mask layout each time. For example, a makeup artist would need eyelid control for visualizing makeup effects, while a plastic surgeon would need  detailed nasal ala editing for the previsualization of a surgery. Achieving this with the previous methods either requires an extensive dataset segmented with various layouts or, by chance, finding a pretrained segmentation network with the intended layout. Neither of these approaches is an attractive direction.

As a viable alternative, we introduce FFaceNeRF, a new NeRF-based face editing method that can significantly enhance the versatility in 3D-aware face editing by few-shot training. FFaceNeRF utilizes a pretrained segmentation network with a fixed mask at the pretraining process while adding a geometry adapter with feature injection to adapt the geometry decoder to the desired mask structure. With the proposed tri-plane features injection and augmentation strategy, Latent Mixing for Triplane Augmentation (LMTA), FFaceNeRF enables effective manipulation of geometric attributes with as few as 10 training samples, based on the adapted mask layouts. Additionally, during inference, our new overlap-based optimization method enables precise and effective editing of small regions.

%Integrating all the proposed components makes FFaceNeRF, an effective few-shot, mask-based 3D facial editing method capable of editing a 3D face after trained with only 10 data. 

To determine which layers from the latent space are most effective for LMTA while preserving semantic information, we evaluated the capability of each layer and assessed its impact. Through comparison with existing NeRF-based facial editing methods, we demonstrate that our method outperforms others in terms of flexibility, control, and resulting image quality. The main contribution of this paper can be summarized as follows:
\begin{itemize}
\item We introduce FFaceNeRF, a few-shot 3D face editing method that employs a newly proposed geometry adapter with feature injection for efficient few-shot training.
\item We propose an overlap-based optimization process to effectively manage customized masks, even in small regions.
\end{itemize}
%This paper details the development and testing of PersoNeRF, offering insights into its applications and foreshadowing the future trajectory of facial image generation technologies.

%The main contribution of this paper can be summarized as follows:
%\begin{itemize}
%    \item We propose PersoNerf, few-shot segmentation based 3D face editing method which has not been explored.    
%
%    \item By injecting the feature from the tri-plane to pretrained geomete
%    
%    \item We define tri-plane mixing as an effective augmentation. By analyzing the tri-plane mixing, we find the best balancing the trade-off between tri-plane mixing ratio and augmentation effeciveness.
%    
%
%\end{itemize}

\vspace{-2mm}
\section{Related work}
\vspace{-2mm}
\subsection{3D-aware Face Image Synthesis}\vspace{-1mm}
With advancements in style-based generators~\cite{karras2019stylegan, karras2019stylegan2}, NeRF~\cite{mildenhall2021nerf}, and 3D Gaussian Splatting~\cite{kerbl20233d}, 3D-aware face generative models have become increasingly popular. These models typically employ NeRF~\cite{gu2021stylenerf, chan2022efficient, deng2022gram} or 3D Gaussian Splatting~\cite{qian2024gaussianavatars, li2024ggavatar, kirschstein2024gghead, xu2024gaussian} representations to generate high-quality 3D faces, producing realistic facial images with fine geometric consistency. Although the quality of the generated images is high in terms of realism and resolution, the conditional generation capacity and controllability of the methods are limited. Therefore, several methods have been proposed to provide a way to use diverse inputs, including sketches~\cite{jo2023cg}, segmentation masks~\cite{chen2022sem2nerf, zhou2023lc}, and facial attribute parameters~\cite{tang20233dfaceshop,li2023preim3d}, for controllable 3D-aware face generation. Because the focus of these methods is on the conditional generation of diverse faces and not on editing of the input identity, they are not suitable for 3D-aware face editing.

\subsection{3D-aware Face Image Editing}\vspace{-1mm}
To edit face images in a 3D-aware manner, several methods has been proposed~\cite{lan2023self,hou2023infamous,hong2022headnerf,zhuang2022mofanerf}. Unlike point-, sketch-,\\ or text-based methods~\cite{gao2023sketchfacenerf,cheng20243d,li2024instructpix}, mask based methods offer precise control by explicitly defining areas for modification while preserving unedited regions. SofGAN~\cite{chen2022sofgan} uses semantic volumes trained on a 3D segmentation dataset for 3D editable image synthesis while FENeRF~\cite{sun2022fenerf} attempts to edit the local shape and texture in a facial volume via GAN inversion. More related to our work, IDE-3D~\cite{sun2022ide} and NerRFFaceEditing~\cite{jiang2022nerffaceediting} utilize disentangled representations for efficient tri-plane-based 3D generative models~\cite{chan2022efficient} and pretrained segmentation networks~\cite{yu2018bisenet} for high-resolution face editing. While these methods have been successful in global editing, their editing capacities are limited to an employed fixed segmentation layout. In contrast, our method utilizes the disentangled representation and segmentation network during the pretraining stage while enabling segmentation mask adaptation through our geometry adapter, trained with only 10 images. This few-shot training process dramatically improves versatility in face editing, allowing fine control over details such as the pupils, nasal area, and more.

\subsection{Utilizing Features from Generative Models}\vspace{-1mm}
Generative models~\cite{karras2019stylegan, karras2019stylegan2, chan2022efficient, gu2021stylenerf, goodfellow2014generative, ho2020denoising, rombach2022high, saharia2022photorealistic} have produced state-of-the-art results in tasks related to unconditional and conditional image generation. Inspired by their powerful image generation capabilities, many attempts have been made to utilize features from pretrained generative models for downstream tasks, as these features are known to contain richer information than original RGB images~\cite{xu2023open, luo2023diffusion, yun2024stylized, li2022bigdatasetgan,yun2024representative,chi2023datasetnerf, ling2021editgan, zhang2021datasetgan, chong2022jojogan, abdal2021labels4free, baranchuk2021label, tang2023emergent}. However, both utilizing tri-plane features and utilizing features for 3D-aware face editing have not been explored. We first conduct experiments on utilizing tri-plane features to confirm if injecting these features helps training the geometry adapter. Then, we further analyze and demonstrate whether augmenting these features enhances the robustness and effectiveness of the geometry adapter. These approaches collectively enhance our method to achieve versatile few-shot face editing.

%the model performance when training on a small number of data .

\section{Methods}\vspace{-1mm}
\subsection{Pretraining}\vspace{-1mm}

\begin{figure}[h]
  \centering
  % the following command controls the width of the embedded PS file
  % (relative to the width of the current column)
  \hspace{-6mm}
  \includegraphics[width=1.05\linewidth]{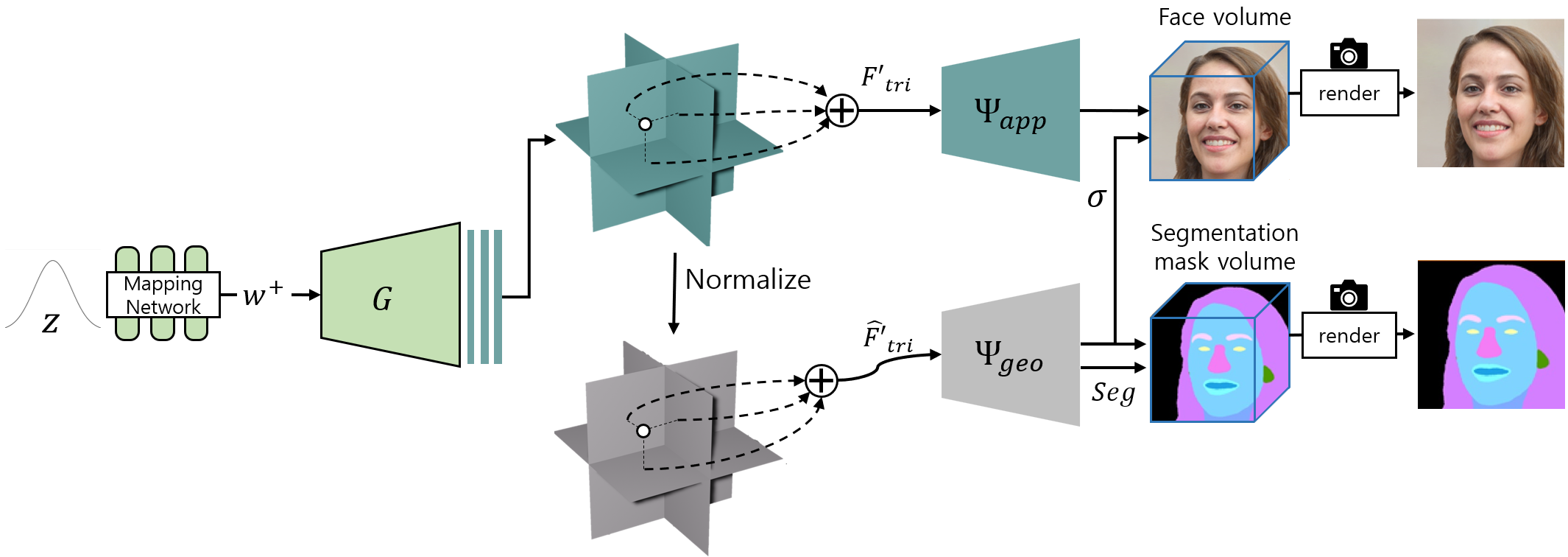}\vspace{-2mm}
    \caption{Pretraining stage of FFaceNeRF following EG3D~\cite{chan2022efficient} and NeRFFaceEditing~\cite{jiang2022nerffaceediting} for disentangled representation.}\vspace{-2mm}
     \label{fig:pretrain}
\end{figure}

FFaceNeRF is built upon EG3D~\cite{chan2022efficient} and follows the pre-training process of NeRFFaceEditing~\cite{jiang2022nerffaceediting} for disentangled appearance and geometry representation as shown in the Figure~\ref{fig:pretrain}. This pre-training starts with training appearance decoder $\Psi_{app}$ that outputs a face volume, which becomes images when rendered. Then, using a pretrained face segmentation network~\cite{yu2018bisenet}, geometry decoder $\Psi_{geo}$ is trained, which outputs a segmentation mask volume that corresponds to the segmentation of the face volume. Here, following NeRFFaceEditing, $\Psi_{app}$ accepts original tri-plane $F'_{tri}$ as input while $\Psi_{geo}$ accepts normalized tri-plane $\hat{F}'_{tri}$. By combining these two trained networks, we can generate a face image and the corresponding segmentation mask from a latent code. Unfortunately, because $\Psi_{geo}$ is trained using a pretrained segmentation network, it can produce a fixed segmentation volume only. 
\begin{figure*}[!t]
  \vspace{-5mm}
  \centering
  % the following command controls the width of the embedded PS file
  % (relative to the width of the current column)
  \includegraphics[width=0.9\linewidth]{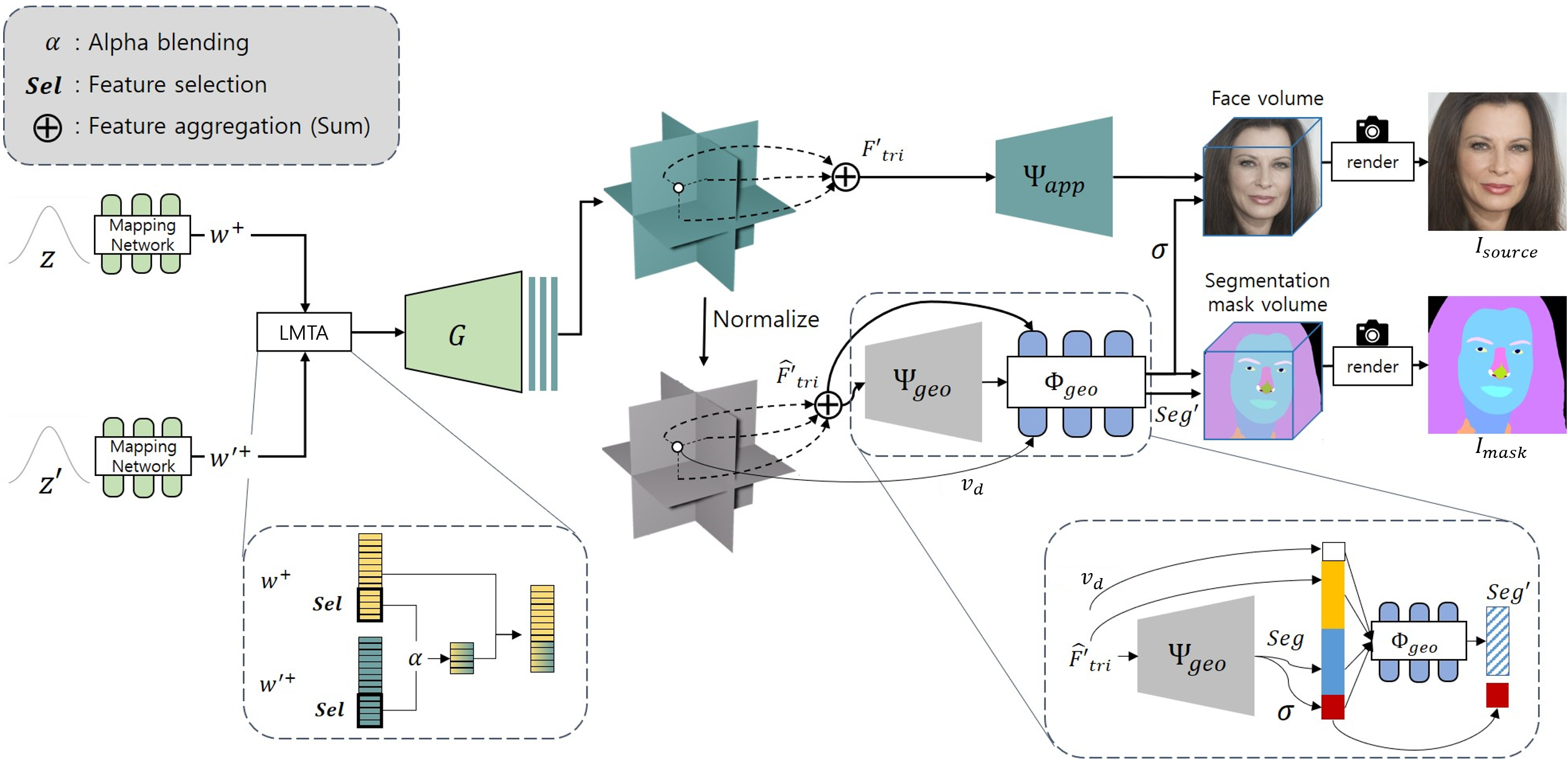}\vspace{-3mm}
    \caption{Overview of FFaceNeRF. LMTA is conducted during the training of $\Phi_{geo}$. $\Phi_{geo}$ takes as input the concatenation of normalized tri-plane feature $\hat{F'}_{tri}$ (yellow box), view direction $v_d$ (white box), outputs of $\Psi_{geo}$, which are segmentation labels $Seg$ (blue box), and density $\sigma$ (red box). Density $\sigma$ is directly used from the output of $\Psi_{geo}$, without further training using $\Phi_{geo}$.}\vspace{-2mm}
     \label{fig:overview}
\end{figure*}

\subsection{FFaceNeRF}\vspace{-1mm}
Our core idea is to add a geometry adapter $\Phi_{geo}$ that modulates the output of $\Psi_{geo}$, producing the desired mask layout. FFaceNeRF can be trained with a small number of data using the process described below.
\begin{enumerate}
\item First, random latent code $w^+$ and corresponding face image $I_{source}$ are generated using a pretrained appearance decoder $\Psi_{app}$ (Top of the Figure~\ref{fig:overview}). With this image, the user makes a few customized segmentation masks in a desired layout.

\item Latent mixing is applied between this $w^+$ and another randomly generated latent $w'^+$ for LMTA, and the geometry adapter $\Phi_{geo}$ is trained with all other parameters frozen.

%\item To edit a target face image at inference, we project the image into the latent space of EG3D and produce the segmentation mask with the trained layout. After the mask is edited by the user, the latent code is optimized according to the mask to produce the edited 3D-aware face image.

\end{enumerate}
In the following subsections, we will describe the structure of geometry adapter $\Phi_{geo}$ with feature injection (Sec.~\ref{subsec:network}), LMTA (Sec.~\ref{subsec:tri-plane}), training process
 (Sec.~\ref{subsec:training}), and finally the proposed editing method (Sec.~\ref{subsec:infer}).

\subsection{Geometry Adapter}~\label{subsec:network}
The geometry adapter $\Phi_{geo}$ is a network that modulates the output of the geometry decoder $\Psi_{geo}$ using a few training data. Because the output dimensions may vary depending on the layout of the mask, $\Phi_{geo}$ is added at the end of $\Psi_{geo}$ to produce a mask volume whose dimensions correspond to the number of layout classes. $\Phi_{geo}$ consists of lightweight MLP, enabling fast training and inference. $\Phi_{geo}$ enables to output the segmentation in a desired layout.

While $\Phi_{geo}$ would work by itself, the pre-training process for $\Psi_{geo}$ focuses on maintaining geometry information for its original classes, causing other information to be discarded. Therefore, we directly inject the tri-plane feature $\hat{F}'_{tri}$ and view direction $v_d$, as shown in the lower right corner of Figure~\ref{fig:overview}. The tri-plane feature, rich in information necessary for face generation, is incorporated before any details are lost. The view direction also guides the geometry information. While NeRF generally does not associate the viewing direction with semantics, most generative NeRFs include data processing steps that align facial features like eyes, nose, and lips. EG3D is one such model where the viewing direction is indeed related to semantics, thus we inject $v_d$ in addition to $\hat{F}'_{tri}$. This information injection and its fusion with the previous output of $\Psi_{geo}$ help $\Phi_{geo}$ to be trained effectively, preserving coarse information while incorporating fine geometric details.

\subsection{Latent Mixing for Triplane Augmentation}~\label{subsec:tri-plane}
The geometry adapter, $\Phi_{geo}$, is trained with a small number of data (e.g., 10 samples) to enable FFaceNeRF to handle diverse segmentation masks. Consequently, avoiding overfitting is a crucial consideration. Therefore, we employ a LMTA, an augmentation which maintains the semantic information while increasing the diversity of the original triplane. Specifically, our back-bone architecture inherits from a style-based generator~\cite{chan2022efficient,karras2019stylegan2} which is known to contain coarse-to-fine information through the early to later layers of the generator. This means that earlier latent code contains more geometric and coarse information while later code contains more color and fine details~\cite{jiang2023nerffacelighting}. When training $\Phi_{geo}$ to adapt the semantic mask volume to a desired layout, details that do not affect the semantic information (such as hue or saturation) can be changed for data augmentation to ensure diverse inputs.
%Because segment-based editing is conducted on geometric part, changing only the fine details that does not change the geometric information can be used as an augmentation.

Additional to the ground-truth latent code $w^+ \in \mathbb{R}^{14 \times 512}$, we randomly sample ${z'}$ and pass it through the pretrained mapping network of EG3D~\cite{chan2022efficient} to generate random latent code ${w'}^+$. This latent code ${w'}^+$ is blended with the $w^+$ using a mixing weight $\alpha$, and the result is passed through the tri-plane generator $G$ to output a modified tri-plane feature ${F'}_{\text{tri}}$. ${F'}_{\text{tri}}$ is directly input to $\Psi_{app}$, while a normalized version, denoted as $\hat{F'}_{\text{tri}}$, is used for both $\Psi_{app}$ and injection to $\Phi_{app}$. For the mixing, only the latent code at a certain selected layer $Sel \in \{0,1\}^{14}$ is used. This $Sel$ is for the latent code whose semantic information is not changed. This process can be written as follows:
\small
\begin{align}\label{eq:alphamixing}
{w'}^+ &= \text{Mapping} ({z'}), \quad\text{where } {z'} \sim \mathcal{N}(0, I)\\ \nonumber
{F'}_{\text{tri}} &= G (\alpha \cdot Sel \cdot {w'}^+ + (1-\alpha) \cdot Sel \cdot w^+ + (1 - Sel) \cdot w^+)
%\label{eq:mixing}
% \\ {F'}_{\text{tri}} &= G (\{ (1 - Sel) \cdot w^+,  (1-\alpha) \cdot Sel \cdot w^+ + \alpha \cdot Sel \cdot {w'}^+  \})
\end{align}\normalsize
The experiments to find the best $Sel$ to choose will be discussed in Sec.~\ref{sec:tri-plane-exp}. %and Sec.~\ref{sec:ablation}.

\subsection{Training}~\label{subsec:training}
FFaceNeRF is designed to facilitate editing of face images after training with a small number of data, as we mentioned in Sec.~\ref{subsec:network}. During training, only the geometry adapter $\Phi_{geo}$ is updated, while other components such as $G$, $\Psi_{app}$, and $\Psi_{geo}$ remain frozen. We train $\Phi_{geo}$ with an augmented tri-plane feature $\hat{F}_{\text{tri}}$ to ensure that the geometry adapter is trained on varied data, enhancing its robustness and generalization. Additionally, we employ regularization to ensure that the ground-truth mask can be generated from non-augmented tri-plane features $F_{\text{tri}}$.

The training objectives consist of two parts, $\mathcal{L}_{CE}$ and $\mathcal{L}_{ovlp}$. $\mathcal{L}_{CE}$ represents a cross-entropy loss, which measures the dissimilarity between the predicted probability $p(y_i)$ and the ground truth class value $y_i$ for pixel $i$. This loss is crucial for ensuring that $\Phi_{geo}$ learns to generate masks that closely match the ground truth. However, $\mathcal{L}_{CE}$ calculates the pixel-wise correct rate which are prone to ignoring classes with small regions. To address this, we adopt an overlap loss $\mathcal{L}_{ovlp}$ using DICE coefficient~\cite{sudre2017generalised}. Because $\mathcal{L}_{CE}$ is faster to optimize, only $\mathcal{L}_{CE}$ is trained first and then $\mathcal{L}_{ovlp}$ is added later. The objective is defined as follows:
\small
\begin{align}
&\mathcal{L}_{total} = \mathcal{L}_{CE} + \lambda \mathcal{L}_{ovlp},\quad \text{where}\label{eq:total_loss} \\
&\mathcal{L}_{CE} = -\sum_{i} \left[ y_i \log(p(y_i)) + (1 - y_i) \log(1 - p(y_i)) \right] \label{eq:ce}\\
&\mathcal{L}_{ovlp} = \left(1 - \frac{2 \sum_{i} p(y_i) y_i + \epsilon}{\sum_{i} p(y_i) + \sum_{i} y_i + \epsilon}\right)
\label{eq:ovlp}
\end{align}\normalsize
Here, $\mathcal{L}_{ovlp}$ quantifies the negative portion of the correct overlaps for each labels and $\epsilon$ is a smoothing factor used to prevent division by zero. $\lambda$ is the weight for the overlap loss using DICE coefficient, which is set to zero during the initial stage.

\vspace{-1mm}
\subsection{Inference}~\label{subsec:infer}
To edit a real face image $I$, we first invert the image $I$ into the latent space $W$ using pivotal tuning inversion~\cite{roich2022pivotal}.
This process yields a projected latent code $w \in W$ and a finetuned generator $G^*$.
With this $w$, FFaceNeRF predicts a segmentation mask $S$ based on the trained data.
The user then edits the mask to produce an edited mask $\bar{S}$.
To generate an edited face image $I'$, we optimize an editing vector $\delta w^+$ so that the predicted segmentation mask $S' \in \mathbb{R}^{C\times W\times H}$ matches the target mask $\bar{S}\in \mathbb{R}^{C\times W\times H}$. Here, $C$, $W$, and $H$ indicate the number of segmentation channel, width, and height, respectively. From this optimization, we obtain a tri-plane feature $F'_{tri}$ that can generate the edited image $I'$.
The process is defined as follows:
\begin{align}   
(S',I') &= (\Phi_{geo}(\Psi_{geo}(\hat{F}'_{tri})),\Psi_{app}(F'_{tri})), \\ \text{where}&\quad F'_{tri} = G^*(w + \delta w^+) \nonumber
\end{align}
The optimization employs the following losses as the minimization targets:
\begin{align}\label{eq:optimize}   
\mathcal{L}_{edit} &= \mathcal{L}_{LPIPS}(I' \otimes (1-r), I \otimes(1-r) ) + \\
&\lambda_{CE}\mathcal{L}_{CE}(S',\bar{S}) + \lambda_{ovlp}\mathcal{L}_{ovlp}(S',\bar{S}), \nonumber
\end{align}
where $\mathcal{L}_{LPIPS}$ indicates an LPIPS~\cite{zhang2018unreasonable} loss whose purpose is to retain the unchanged region $(1-r) \in \mathbb{R}^{1\times W\times H}$. Both $\mathcal{L}_{CE}$ and $\mathcal{L}_{ovlp}$ ensure editing fidelity. This optimization process differs from the conventional probability-based approaches~\cite{ling2021editgan,jiang2022nerffaceediting}. The conventional approaches compare the cross-entropy between $S'$ and $\bar{S}$, and therefore often fail to handle changes of labels in small regions. Our overlap-based optimization process ensures that $S'$ is aligned with $\bar{S}$ even for small regions while preserving the rest of the image unchanged.
%{-0.2cm}
\section{Experiments}\vspace{-1mm}
\subsection{Implementation details}~\label{sec:implement}
We trained and tested our FFaceNeRF on a computer with an NVIDIA RTX A6000 GPU. The training of $\Phi_{geo}$ was conducted for 5,000 steps with the batch size of 4. The $\lambda$ in Eq.~\eqref{eq:total_loss} was set to 0 for the first 4,000 steps and set to 0.1 for the last 1,000 steps, and $\lambda_{CE}$ and $\lambda_{ovlp}$ used in Eq.~\eqref{eq:optimize} were set to 0.5 and 1, respectively. The mixing rate $\alpha$ used in Eq.~\eqref{eq:alphamixing} was set to 0.5 in all of our experiments. The learning rate was increased until 0.03 and then decreased until 0 at the end of the training using OnecycleLR~\cite{smith2019super}. The total training required around 40 minutes.

\subsection{Dataset}\vspace{-1mm}
\begin{figure}[t]
  \centering
  \vspace{-3mm}
  \includegraphics[width=0.9\linewidth]{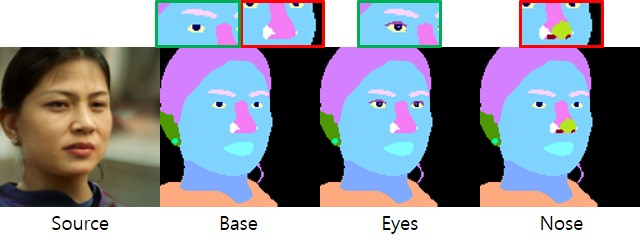}
    \vspace{-10px}
    \caption{Examples of dataset with different segmentation layouts. Green boxes are close-up views of eye regions while red boxes are close-up views of nose regions.}
    \vspace{-10px}
     \label{fig:dataset}
\end{figure}

%Examples of dataset with different segmentation layouts.

For experiments, we prepared three different mask datasets with different layouts: Base segmentation layout (Base), Eyes-specialized segmentation layout (Eyes), and Nose-specialized segmentation layout (Nose). Base consisted of 17 classes, while Eyes and Nose were composed of 19 classes each. The Eyes and Nose layouts were built upon the Base layout by adding details for each layout as shown in Figure~\ref{fig:dataset}. For training, we annotated 10 segmentation masks from randomly generated images, sharing the source identity across all three datasets. For quantitative evaluation, we annotated 22 segmentation masks in the Base layout. Another 30 segmentation masks in the Base layout were annotated for an additional dataset scaling experiment. For qualitative evaluation, we randomly sampled 40 images from CelebA-HQ~\cite{karras2017progressive} and edited the mask using Eyes and Nose layouts.

\begin{figure}[t]
  \centering
  % the following command controls the width of the embedded PS file
  % (relative to the width of the current column)
  \includegraphics[width=\linewidth]{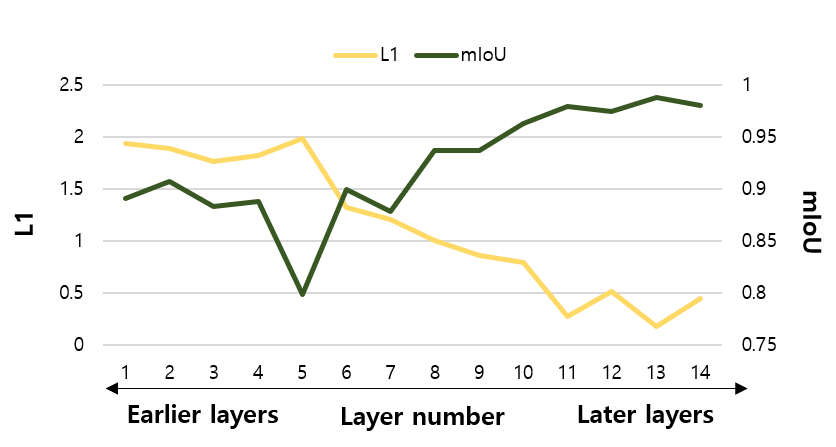}
  \vspace{-20px}
  \caption{Semantics-augmentation tradeoff: When mixing earlier layers, semantics and tri-plane feature information change largely (high L1, low mIoU). On the other hand, when mixing later layers, semantics and augmentation change little (low L1, high mIoU).
  % : Equivariant spatial augmentation and invariant magnitude augmentation.
  }
  \vspace{-10px}
     \label{fig:augmentation}
\end{figure}
\subsection{Latent Mixing for Triplane Augmentation}\vspace{-1mm}~\label{sec:tri-plane-exp}
The style-based generator is known to contain coarse to fine details as the layers progress outwards. We conducted an experiment on mixing a single layer with the latent code of $w^+_i$, where $i \in [1, 14]$ indicates the index of layer. This experiment was conducted to determine in which layer the semantic and geometric information changes or remains unchanged. We randomly sampled 1,000 latent codes $w^+$ from random $z$ passed through the mapping network and generated corresponding images $I$. In addition, we sampled another $\bar{w}^+$ from random $\bar{z}$. We mixed this latent code $\bar{w}^+$ with $w^+$ only in the $i$-th layer and generated an image $\bar{I}$. The mean Intersection over Union (mIoU) was calculated between $I$ and $\bar{I}$ using their segmentation masks obtained from the pretrained segmentation network~\cite{yu2018bisenet}, to determine until which layer we can utilize LMTA. A high mIoU indicates the retention of semantic information after the mixing.

In addition to mIoU, we calculated the L1 distance between the tri-plane $G(Sel \cdot \bar{w}^+ + (1-Sel) \cdot w^+)$ and $G(w^+)$ to check the augmentation amount. A higher L1 value indicates a larger difference between the augmented tri-plane and the original tri-plane, allowing the network to be trained with more diverse data. As shown in Figure~\ref{fig:augmentation}, we observed a trade-off between semantics preservation and augmentation effectiveness depending on the selection of mixing layers. When earlier layers are selected, semantics change, failing to maintain the geometric structure of the original image, but diversity increases. Conversely, selecting later layers results in higher retention but modest diversity.

%When later layers are selected, semantics do not change, preserving the overall structure of the image. Conversely, selecting earlier layers enhances the effectiveness of augmentation

\begin{figure}[t]
  \centering
  \vspace{-3mm}
  \includegraphics[width=\linewidth]{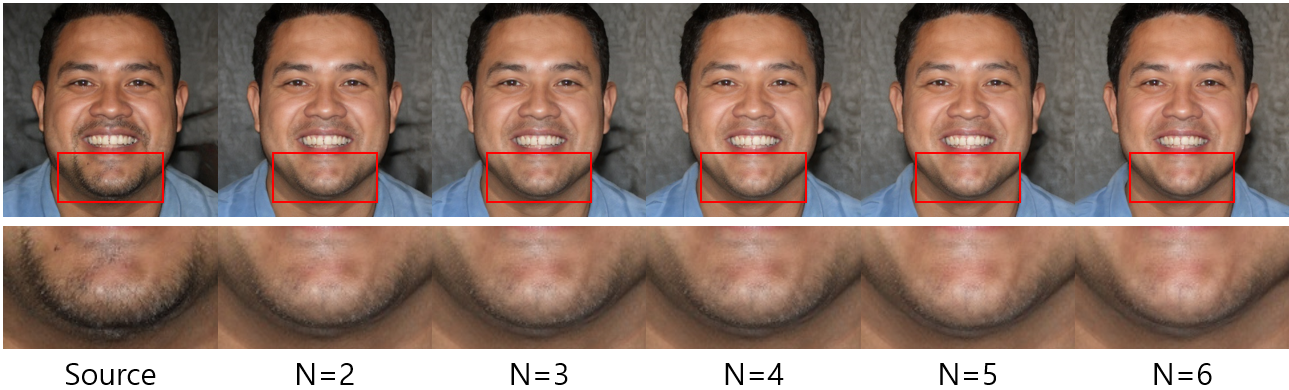}\vspace{-1mm}
  \caption{Visualization of tri-plane mixing N layers with the top L1 metrics. The geometry and the semantic information such as the outline of chin and beard are changed even when only 2 layers are mixed.}
   \label{fig:mixb}
\end{figure}
\begin{figure}[t]
  \centering
  \includegraphics[width=\linewidth]{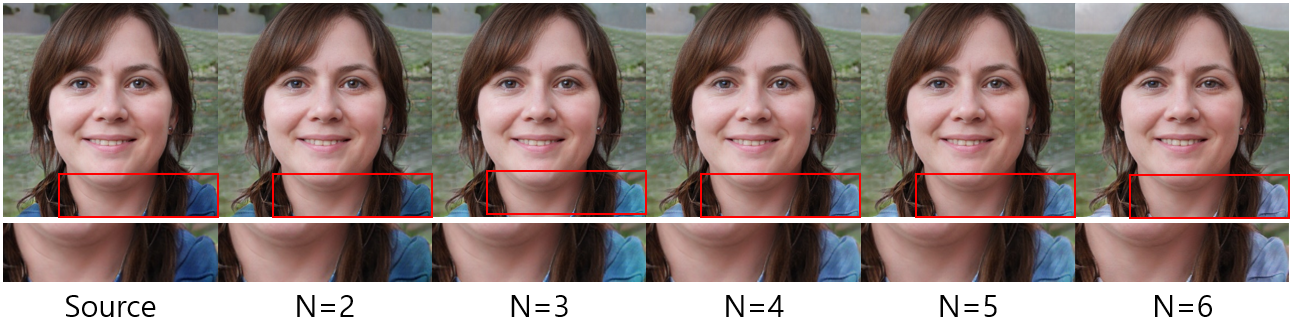}\vspace{-1mm}
  \caption{Visualization of tri-plane mixing N layers based on the top mIoU metrics. The geometry and the semantic information are not changed while colors are changed.}
  \vspace{-2mm}
   \label{fig:mixa}
\end{figure}

\begin{figure}[t]
  \hspace{-4mm}
  \vspace{-3mm}
    \includegraphics[width=1.08\linewidth]{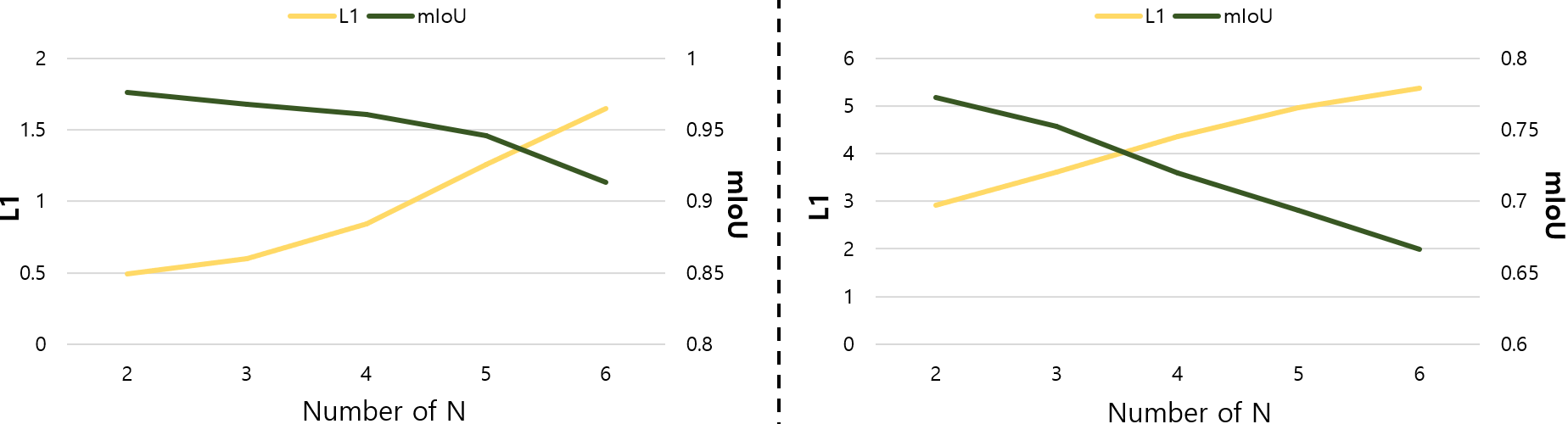}\vspace{-1mm}
  \caption{Results of mixing multiple tri-planes. Left: Mixing top N layers based on mIoU, Right: Mixing top N layers based on L1.}
     \label{fig:mixing}
\end{figure}

Additionally, we conducted experiments on mixing multiple layers to find better augmentation effect.
Specifically, we tested with mixing the top N layers and measured mIoU and L1 within the range of N=[2, 6]. For example, for N=2 with top mIoU, we chose layers 13 and 14, producing the highest and the second highest mIoU values, as shown in Figure~\ref{fig:augmentation}. For N=3 with top L1, we mixed layers 5, 1, and 2. We listed the top 6 layers in mIoU and L1 in Table 1 of supplementary material. 
%~\ref{tab:top6_layers}.
%For instance, referring to Table~\ref{tab:top6_layers}, when N=2 for mIoU, we mixed layers 13 and 14, and when N=3 for L1, we mixed layers 5, 1, and 2.
We measured mIoU and L1 for these configurations, and the results are shown in Figure~\ref{fig:mixing}.
In both cases, we observed that as the number of mixed layers increased, mIoU decreased while L1 increased.
However, in the case of mixing the top N layers based on L1 (Figure~\ref{fig:mixing}, right), the mIoU values were consistently very low, less than 0.8 for all N.
This indicates that mixing layers based on a high L1 value is not suitable for local editing tasks.
As shown in Figure~\ref{fig:mixb}, even mixing just two layers (N=2) resulted in changes to the outline of the chin.

On the other hand, for mixing the top N layers based on mIoU (Figure~\ref{fig:mixing}, left), as N increased, the L1 value consistently increased, while the mIoU value did not show significant changes until N=5. Additionally, the L1 and mIoU values were relatively higher overall compared to mixing the top N layers based on L1.
Based on these results, we opted to mix the latent codes of the top 5 layers based on mIoU (layers $\in$ [10, 14]) for our method.
As shown in Figure~\ref{fig:mixa}, the geometry or semantic information of the source does not change even when 5 layers are mixed, while only the hue of the clothes and the brightness of the face change.

%\begin{figure}[h]
%\hspace{-6mm}\makebox[\textwidth][l]{%
%    \includegraphics[width=1.1\linewidth]{figures/ours.png}}
%\caption{Examples of multi-view images edited using our method. Edited regions are indicated with red arrows in target masks.}\label{fig:results}
%\end{figure}

\begin{figure}[t]
    \hspace{-5mm}
    \includegraphics[width=1.1\linewidth]{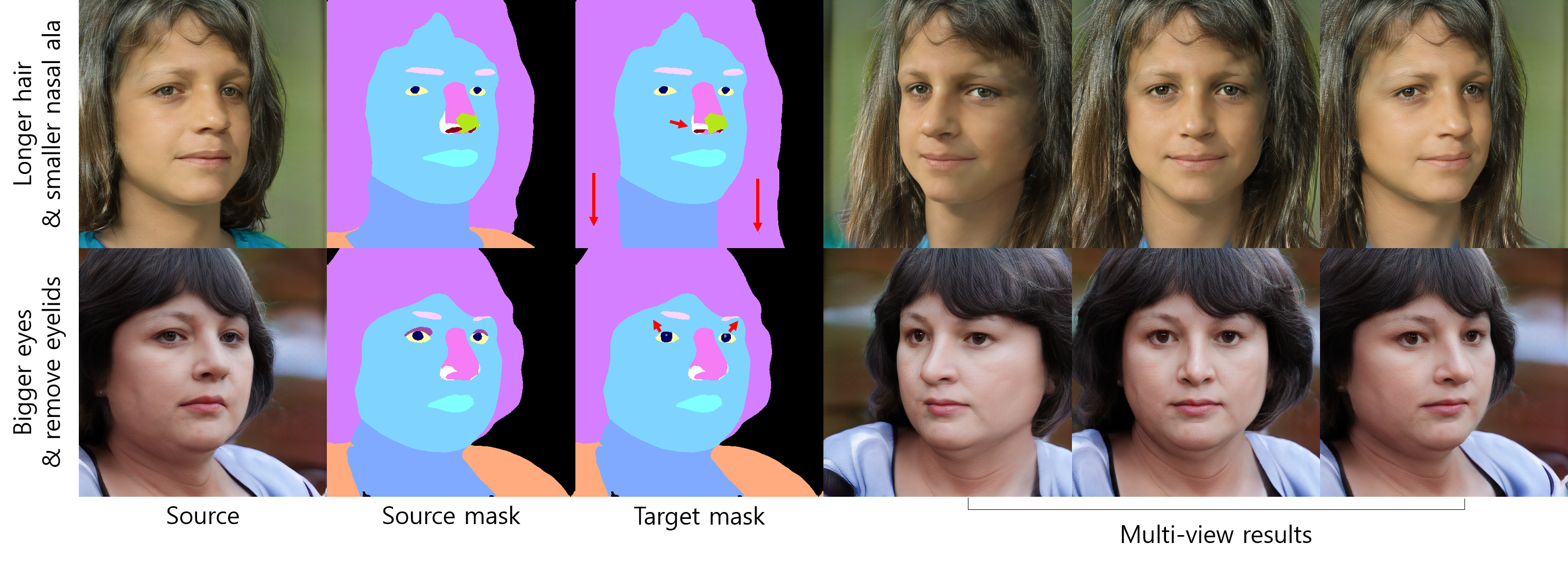}\vspace{-3mm}
\caption{Examples of multi-view images edited using our method. Edited regions are indicated with red arrows in target masks.}\label{fig:results}
\vspace{-2mm}
\end{figure}

\subsection{Evaluations}~\label{sec:evaluation}
\vspace{-7mm}
\subsubsection{Qualitative Results}
We evaluated our method based on two main features: editing capability and multi-view functionality. The results are displayed in Figure~\ref{fig:results}, which were edited from test-set identities unseen during training. The results demonstrate that our method can produce intended outcomes following the edited segmentation mask. Additionally, when rendered from different views, the edited face regions are well-blended with the original face structure. This indicates that our adaptation method effectively learns to follow the mask layout when editing a face image while preserving 3D information from few-shot training.

\begin{figure*}[t]
  \centering
\vspace{-5mm}
  \includegraphics[width=0.88\linewidth]{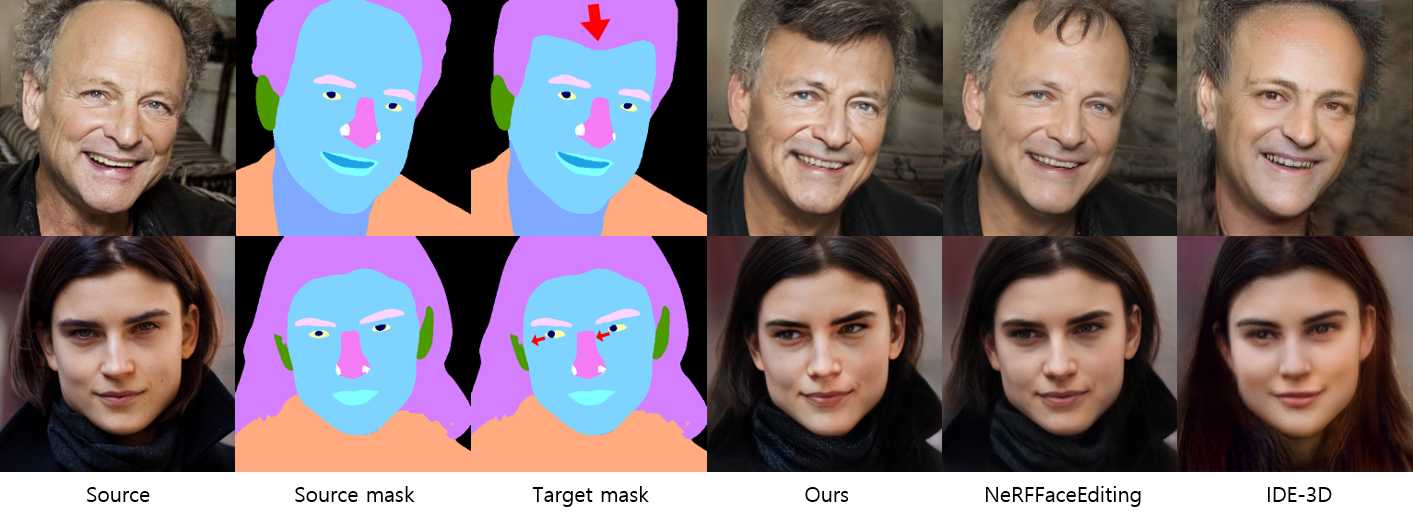}
          \vspace{-4mm}
        \caption{Examples of our results and those of baseline methods in editing tasks. The results of our method faithfully reflect the edited regions.}
        \label{fig:comp}
\end{figure*}\vspace{-2mm}

\subsubsection{Baseline Comparisons}\vspace{-1mm}
%While there are no existing few-shot 3D-aware face editing methods with masks, we chose the nearest competitors. The most similar to ours is NeRFFaceEditing~\cite{jiang2022nerffaceediting}, which is also our backbone model. It is a 3D-aware face editing method with disentangled tri-plane representation. IDE-3D~\cite{sun2022ide} is another method using 3D-aware face editing with tri-plane representation. NeRFFaceEditing and IDE-3D, which are trained with more than thousands of samples with a segmentation network, cannot be trained with just 10 examples as our method can. Therefore, we only trained the geometry decoder of each model while keeping other components fixed for a fair comparison. We did not compare with SofGAN~\cite{chen2022sofgan} because it needs to train the occupancy field and a conditional GAN, which cannot converge with 10 images, nor fine-tune semantic occupancy field because it is then input to a conditional GAN.

To evaluate the editing performance of our method, we compared ours with other 3D-aware face editing methods. We chose two state of the art mask based 3D-aware face editing methods: NeRFFaceEditing, which serves as our backbone model, and IDE-3D, which performs similar editing tasks using tri-plane representation.
Unlike our method, those methods typically require more than a few thousand samples along with a segmentation network to train other segmentation masks. Therefore, for a fair comparison, we conducted transfer learning by training only the geometry decoder of each model with our same dataset while keeping all other components frozen. This ensured that the comparison focused solely on the effectiveness of the geometry decoder in the context of few-shot learning. %For this comparison, we prepared 40 image samples randomly selected from CelebA-HQ~\cite{karras2017progressive} which was not used in our training and constructed three segmentation masks (Base, Eyes, and Nose) for each image.

We present qualitative comparison results in Figure~\ref{fig:comp}, visualizing the editing results of our method, NeRFFaceEditing, and IDE-3D. %For the comparison on real images, we randomly selected 40 images from CelebA-HQ~\cite{karras2017progressive}. 
As shown in the figure, our results reflected the target mask most faithfully, while NeRFFaceEditing reflected it to only a small extent. IDE-3D did not produce correctly edited results and often failed to preserve the original identity of the person. This indicates that our method can perform editing better than the comparative methods in terms of following the edited target mask.

To compare visual quality in human perception, we conducted a perceptual study comparing the results of each method given edited target segmentation masks. We evaluated the editing performance of our method compared to NeRFFaceEditing and IDE-3D for editing using an A/B questionnaire. Participants were asked to choose the method that better satisfied the following three criteria: faithfulness to the changed regions, retention of the unchanged regions, and overall visual quality. For comparison, we presented the original image, semantic mask, modified semantic mask, and two results in a random order. Each participant was asked to answer 90 questions from 30 comparisons (30 ours vs. 15 others each), with a "no difference" option to avoid random guessing.

A total of 21 participants (aged from 23 to 32, all with normal vision) were recruited for the study. The percentages of A/B testing results are shown in Table~\ref{Tab:user}. The percentage indicates the proportion of participants choosing our method over the competitor ($ours\over{ours +  competitor}$$\times 100(\%)$). In all questions, ours achieved more than 50\%, indicating that participants rated our method higher in faithfulness, source retention, and visual quality compared to both competitors. %Meanwhile, NeRFFaceEditing scored close to ours in retention.

\begin{table}[t]
\centering
\setlength{\tabcolsep}{1mm}
\caption{Perceptual study results on face editing. The number indicates the percentage of selections for our methods over competitors.}\vspace{-5mm}\small
\begin{tabular}{|l|c|c|c|}
\noalign{\smallskip}\noalign{\smallskip}\hline
Comparison  & Faithfulness(\%)& Retention(\%) & Quality(\%)  \\
\hline
vs NeRFFaceEditing     & 72.29 & 67.83 & 68.68   \\ 
vs IDE-3D  & 79.65  & 80.17 & 81.22 \\
\hline
\end{tabular}
\label{Tab:user}
\end{table}

\begin{table}[t]
\centering
\setlength{\tabcolsep}{5mm}
\caption{Quantitative results of our method and NeRFFaceEditing on mask generation. Highest scores are denoted in bold.}\small\vspace{-5mm}
\begin{tabular}{|l|c|}
\noalign{\smallskip}\noalign{\smallskip}\hline
Methods  & Average mIoU [min, max] (\%) \\
\hline
Ours                   & \textbf{85.33  [84.8, 85.7]}\\ 
NeRFFaceEditing  & 81.37 [81.2, 81.5]           
\\
\hline
\end{tabular}
\vspace{-4mm}
\label{Tab:addquant}
\end{table}
% \vspace{-4mm}

We additionally performed a quantitative comparison with NeRFFaceEditing on a mask generation task. Faithfully generating a mask is the first step toward effective editing, a process that our editing method shares with NeRFFaceEditing. The mask generation task was conducted 5 times to reduce the effect of randomness, and the average, minimum, and maximum values of the 5 trials were noted in Table~\ref{Tab:addquant}. Our method performed better than the comparative method in generating correct masks using 22 Base layout test sets. We attribute this superior performance to our adaptation method.

\vspace{-2mm}
\subsubsection{Ablation Study}~\label{sec:ablation}
\begin{figure}[t]
  \centering
  \hspace{-2mm}
  \includegraphics[width=1.05\linewidth]{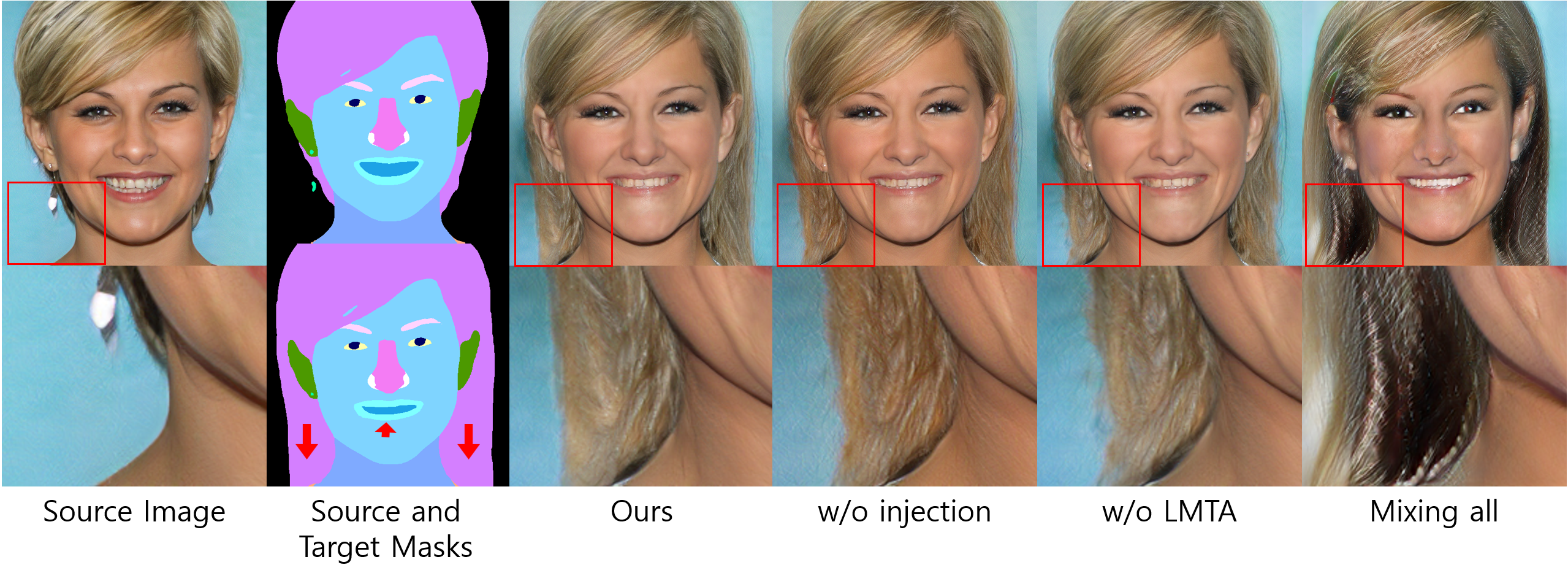}
        \vspace{-6mm}
        \caption{Ablation study results. Our full model best followed the target masks while preserving the original identity. The color shifted when trained without feature injection, hair was not faithfully generated when trained without LMTA, and the source identity changed when trained with mixing all layers. }
        %especially on the eyes and nose.}
        \label{fig:ablation}
        \vspace{-2mm}
\end{figure}

To perform an ablation study, we trained our model using different settings to identify the effectiveness achieved by each of our components: 1) Without feature injection; 2) Without augmentation (LMTA); 3) Using all layers for mixing; 4) our full model. We tested with different numbers of data to evaluate the effects of each component with varying data sizes. For this evaluation, mIoU was used for 22 test sets. As shown in the Table~\ref{Tab:ablation}, a method without injection resulted poor even with a larger training data such as 30 sets, a method without LMTA suffered when a small number of training data was used due to overfitting, and mixing all layers resulted worst due to semantic changes during its strong augmentation. On the other hand, our full method produced the best results across 5 to 30 training data. For training with one data, ours record the second-best results because it is too small to adapt to a new mask layout. This is also shown in the visualization presented in Figure~\ref{fig:ablation} which was produced by all methods trained with five data. The results from our methods did not show color shift while following the target mask.

\begin{figure}[t]
\centering
    \includegraphics[width=1\linewidth]{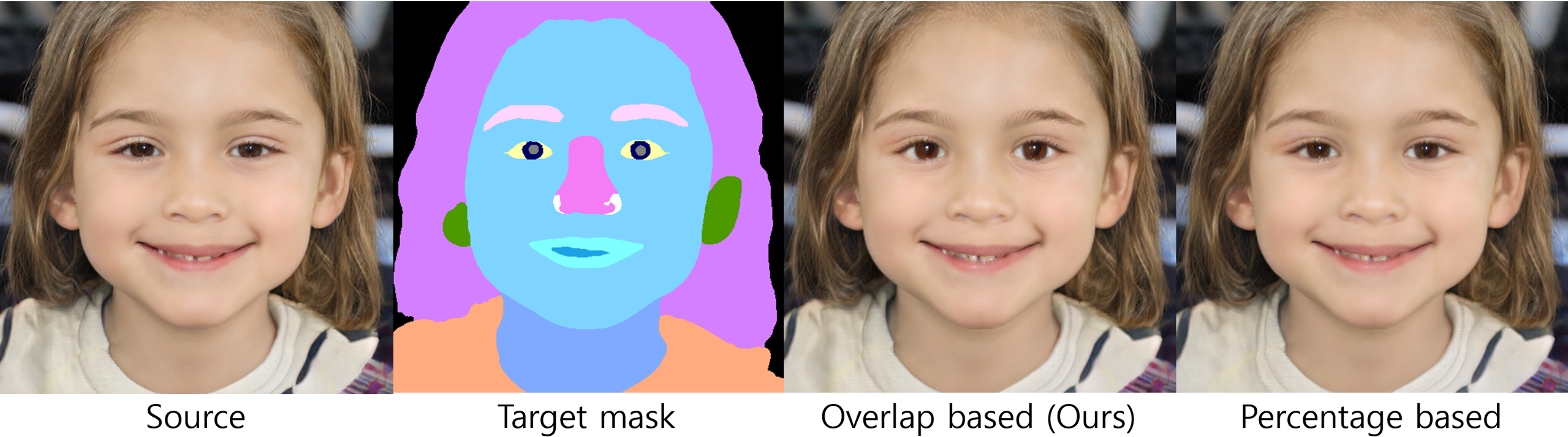}\vspace{-3mm}
\caption{Comparisons with percentage-based optimization on enlarging eyes show that our method more faithfully follows the desired eye size modifications.}\label{fig:addab}
\end{figure}\vspace{3mm}

\begin{table}[t]
\centering
\renewcommand{\arraystretch}{1.2}
\setlength{\tabcolsep}{2mm}
\caption{Quantitative results of ablation study on mask generation with a different number of training data. The highest scores are denoted in bold.}\vspace{-5mm}\small
\begin{tabular}{|l|c|c|c|c|c|c|}
\noalign{\smallskip}\noalign{\smallskip}\hline
Number of Data & 1 & 5 & 10 & 20 & 30 \\
\hline
Ours                  & 0.711 & \textbf{0.832} & \textbf{0.850} & \textbf{0.855} & \textbf{0.860} \\ 
w/o injection & \textbf{0.741} & 0.806 & 0.835 & 0.844 & 0.847 \\ 
w/o LMTA      & 0.695 & 0.829 & 0.845 & \textbf{0.855} & 0.859 \\
Mixing all     & 0.603 & 0.654 & 0.743 & 0.785 & 0.780 \\
\hline
\end{tabular}
\vspace{-4mm}
\label{Tab:ablation}
\end{table}

\begin{figure*}
\vspace{-5mm}
\makebox[\textwidth][c]{
  \includegraphics[width=0.85\linewidth]{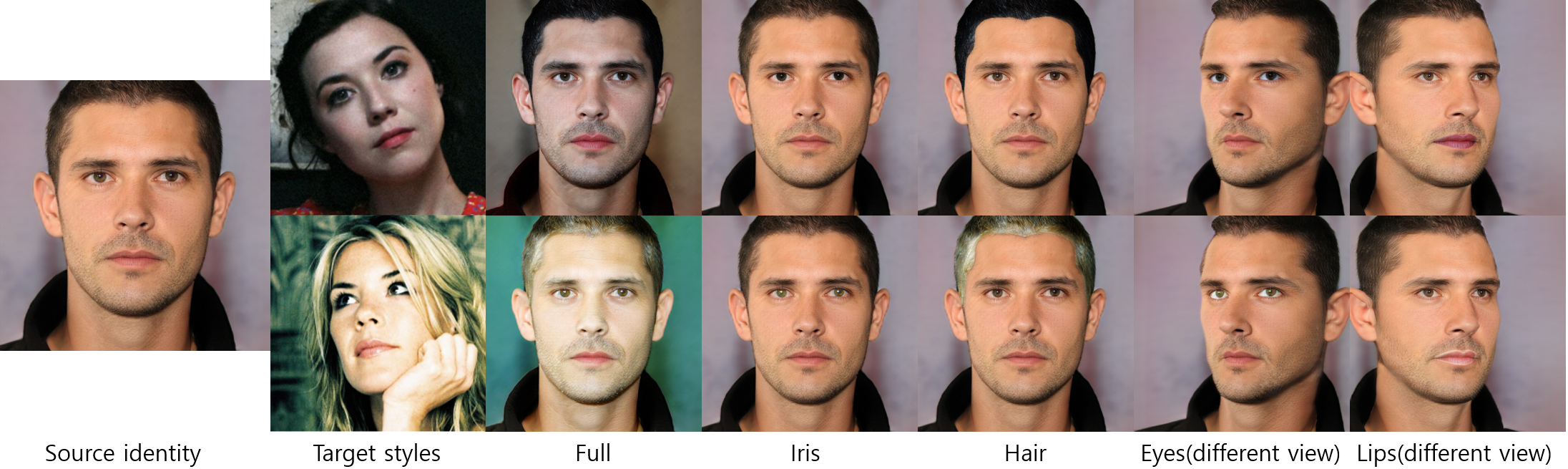}}
  \vspace{-6mm}
\caption{Results of full and partial style transfer application.}\label{fig:application} %Source images closely follow the style images within the given mask label or across the entire image.}
\vspace{-2mm}
\end{figure*}

We further conducted a visual comparison between the results of our overlap-based optimization and the previous percentage-based optimization~\cite{ling2021editgan,jiang2022nerffaceediting}. As shown in Figure~\ref{fig:addab}, our overlap-based optimization produced bigger eyes, following the given layout. This indicates that overlap based optimization is crucial for small segments, such as eye editing.
%\begin{table}[t]
%\centering
%\setlength{\tabcolsep}{3mm}
%\caption{Quantitative results of ablation study on mask generation. Highest scores are denoted in bold.} %{-10pt}
%%\vspace{px}
%\begin{tabular}{|l|c|}
%\noalign{\smallskip}\noalign{\smallskip}\hline
%Methods  & Average mIoU [min, max] (\%)\\
%\hline
%Ours                   & \textbf{85.33  [84.8, 85.7]}\\ 
%w/o feature injection  & 83.76  [83.4, 84.2]     \\
%Single stage           & 84.71  [84.1, 85.0]     \\
%w/o augmentation       & 84.71  [84.4, 85.0]     \\
%%Mixing one             & 84.27  [83.8, 84.7]     \\
%mixing all             & 74.60  [74.3, 74.8]      \\
%\hline
%\end{tabular}
%\vspace{-5px}
%\label{Tab:ablation}
%\end{table}

\vspace{-1mm}
\section{Application}
\begin{figure}
  \centering
  \includegraphics[width=\linewidth]{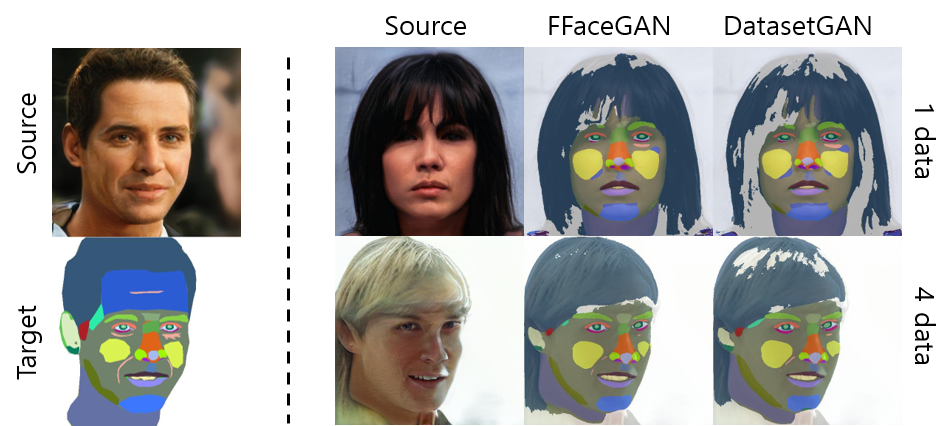}
  \vspace{-5mm}
\caption{Comparison between our FFaceGAN and DatasetGAN in generating segmentation masks based on the target labels.}\label{fig:FFaceGAN}  \vspace{-2mm}
\end{figure}

\vspace{-1mm}
\paragraph{Partial Style Transfer}
Tri-plane-based NeRF offers a powerful framework for facial representation, and style transfer is one of the key applications for face editing. FFaceNeRF leverages disentangled geometry and appearance representations to modify facial style while preserving geometry. Additionally, use of a customized mask makes this editing possible with desired partial labels. To produce the results, the mean and variance of the tri-plane from the target style images were extracted, and the normalized source tri-plane was denormalized using these values. This denormalized tri-plane feature was then passed through the appearance decoder, $\Psi_{app}$, for full style transfer. Afterward, using the generated masks in the desired layout, only the selected label was blended with the original face. To ensure a seamless transition without visible boundaries, linear blending was applied at the edges. As shown in Figure~\ref{fig:application}, our method successfully transfer the style for the desired regions.
\vspace{-2mm}
\paragraph{FFaceGAN} 
DatasetGAN~\cite{zhang2021datasetgan} is a powerful framework that generates segmentation using limited data with a pretrained StyleGAN~\cite{karras2019stylegan}. To demonstrate the effectiveness of our adapter and LMTA, we integrated these into DatasetGAN, resulting in FFaceGAN. A comparison between DatasetGAN and our FFaceGAN is presented in Figure~\ref{fig:FFaceGAN}. The results of FFaceGAN show dramatic improvement of the quality over those of DatasetGAN. For example, artifacts such as segmentation holes and unlabeled regions were reduced, illustrating the effectiveness of our proposed strategies. FFaceGAN demonstrates that the proposed techniques can be applied to architectures beyond NeRFFaceEditing and EG3D. Additional details and explanations for both applications are provided in the supplementary material.

%This is made possible by our training process, which incorporates the disentangled representation without changing the appearance decoder. For partial style transfer, a mask label is selected for the source image

%To achieve this, the mean and variance of the tri-plane from the style images are extracted, and the normalized source tri-plane is denormalized using these values. This denormalized tri-plane feature is then passed through the appearance decoder, $\Psi_{app}$. This is made possible by our training process, which incorporates the disentangled representation without changing the appearance decoder. For partial style transfer, a mask label is selected for the source image

\section{Limitation and Conclusion}

In this paper, we introduced FFaceNeRF, a novel NeRF-based face editing technique that enhances mask-based 3D-aware face editing through few-shot training. FFaceNeRF overcomes limitations of existing methods that often depend on extensive datasets or pretrained segmentation networks, by significantly reducing the number of required training samples. This approach offers greater user control and customization.

To achieve this, we proposed a geometry adaptor with feature injection, which enables effective manipulation of geometric attributes using a small number of training data. Additionally, we adopted LMTA that preserves semantic information while improving the efficiency of data augmentation. To further refine editing capabilities on small labels, we introduced an overlap-based optimization technique. Finally, we demonstrated the potential for broader applications beyond existing frameworks like EG3D and NeRFFaceEditing.

While FFaceNeRF demonstrated strong performance in few-shot segmentation-based editing, achieving real-time performance remains challenging. Our inferences are conducted with iterative optimization, requiring around 31 seconds per edit. One potential way to overcome this limitation is to train an encoder for customized masks. However, training an encoder with as few as 10 data samples has not been well-researched. Therefore, utilizing a pretrained mask encoder to develop customized encoder for interactive face editing could be a meaningful future research direction. In addition, performance in one-shot setting is limited. Because FFaceNeRF requires to train a geometry adapter, it cannot generalize to all faces when only one data is provided even with our augmentation. Incorporating SAM-like models~\cite{kirillov2023segment} could help facilitate a one-shot approach due to their general knowledge of segmentation.

\clearpage
\section*{Acknowledgements}
This work was supported by Institute of Information \& Communications Technology Planning \& Evaluation (IITP) grant funded by the Korea government (MSIT) (No.RS-2024-00439499).

%While FFaceNeRF demonstrated strong performance in few-shot segmentation-based editing, achieving real-time performance remains challenging. Our inferences are conducted with iterative optimization, requiring around 31 seconds per edit.
%\input{sec/99_suppl.tex}

{
    \small
    \bibliographystyle{ieeenat_fullname}
    \bibliography{main}
}


\end{document}